\documentclass{article}
\usepackage{pdfpages}
\usepackage{xcolor}
\usepackage[mathlines]{lineno}
\usepackage{graphicx}
\usepackage{cite}
\usepackage{caption}
 
\usepackage{subfig} 
\usepackage{subcaption}
\usepackage{authblk}
\usepackage{amsmath}
\usepackage[top=5cm, bottom=5cm, left=3cm, right=2cm, headsep=1.5cm, heightrounded=true]{geometry}
\usepackage{fancyhdr}
\graphicspath{ {{images/}} }

\title{Holographic Central Charge Effects on Black Hole Thermodynamics and Quantum Information}

\author[1]{Yahya Ladghami\thanks{ \texttt{yahya.ladghami@ump.ac.ma}}}
\author[1,2]{Taoufik Ouali\thanks{ \texttt{t.ouali@ump.ac.ma}}}
\affil[1] {Laboratory of Physics of Matter and Radiation, Mohammed I University, BP 717, Oujda, Morocco}
\affil[2]{Astrophysical and Cosmological Center, BP 717, Oujda, Morocco}

\begin{document}
	\maketitle
\begin{abstract}

In this paper, based on the Anti-de Sitter/Conformal Field Theory (AdS/CFT) correspondence, we highlight the fundamental role of the holographic central charge in connecting the boundary theory to quantum information, black hole thermodynamics, and the nature of gravity in the bulk. We establish that the large central charge of the boundary conformal field theory corresponds to classical gravity, while a small central charge corresponds to quantum gravity described by Loop Quantum Gravity. In addition, we study the thermodynamic behavior of AdS-Schwarzschild black holes for both large and small central charges. For large central charge, the classical AdS-Schwarzschild black holes have two phases: unstable small black holes and stable large black holes. Conversely, for small central charge, black holes are stable, and their entropy is smaller than that of classical black holes. To explore the influence of the boundary central charge on the information loss paradox, we use the island formula to recover the Page curve. We find that before the Page time, the entanglement entropy of Hawking radiation increases with time, and its slope is determined by the central charge of the boundary theory. After the Page time, the island inside black holes emerges, and the unitarity of black hole evaporation is restored, yielding a constant entropy consistent with the Page curve. This entanglement entropy, i.e. after the Page time, depends on the Bekenstein-Hawking entropy and includes a logarithmic correction related to the central charge.

\end{abstract}

\section{Introduction}

\hspace{15pt} Black holes serve as a laboratory for testing and studying the major ideas in theoretical physics, such as quantum gravity, the microstructure of spacetime, the unification of fundamental forces, and the information loss paradox. Black holes are among the most fascinating systems in the universe and play a cornerstone role in uncovering its fundamental mysteries. They are not only gravitational systems characterized by strong gravitational fields, singularities, and event horizons, but they also evaporate through Hawking radiation due to quantum effects \cite{1}. Moreover, black holes possess entropy proportional to their surface area and obey four laws analogous to the laws of thermodynamics \cite{2,3}. These properties make black holes extremely important, as they provide a deep connection between gravity and quantum mechanics. \\

On the other hand, within the framework of the Anti-de Sitter/Conformal Field Theory  correspondence \cite{4}, black holes in Anti-de Sitter spacetime play a fundamental role  as they  correspond to a thermal state in the boundary Conformal Field Theory (CFT), where the Hawking temperature and entropy of the black hole match the temperature and entropy of CFT, respectively. There also exists an equivalence between the thermodynamics of CFT and that of AdS black holes, represented by the holographic dictionary \cite{5}. Based on this correspondence,  holographic description of black hole thermodynamics has been constructed in the literature, often referred to as holographic thermodynamics \cite{6,7,8,A1,A2,A3,A4}. In the AdS/CFT framework, the degrees of freedom of the boundary theory  are encoded in the central charge, which plays a fundamental role in determining the nature of gravity and spacetime in the bulk. A large central charge (corresponding to a large number of degrees of freedom) in CFT represents a weakly coupled bulk gravity theory \cite{4,9,10}, that is, classical gravity as described by general relativity. Conversely, a small central charge in CFT corresponds to a strongly coupled bulk regime \cite{11,12,13,14}, that is, a quantum gravity theory such as Loop Quantum Gravity (LQG) \cite{15}. Thus, the boundary central charge controls whether the bulk spacetime can be described by classical gravity or requires a quantum gravitational framework. In this work, we focus on AdS Schwarzschild black holes and investigate how their physics changes between the classical  and quantum  regimes of gravity, from thermodynamics to the information paradox. \\

The black hole information loss paradox is a fundamental problem in theoretical physics, arising from the fact that black holes evaporate by emitting Hawking radiation \cite{1}. This radiation is thermal, and the process seemingly evolves a pure quantum state into a mixed state, implying a loss of information and violating the unitarity principle of quantum mechanics \cite{16}. To preserve unitarity, the entanglement entropy of Hawking radiation outside the black hole should follow the Page curve that describes the evolution of entropy over time \cite{17}.  One proposed resolution is the firewall approach \cite{AMPS2013}, which suggests that preserving unitarity necessitates the destruction of the smooth event horizon by high-energy zone, known as firewall. In this scenario, the entanglement between interior and exterior Hawking modes is broken after the Page time, thereby preventing information loss at the expense of violating the equivalence principle.
Recently, a new approach, has been proposed to resolve this issue \cite{18}. According to this approach, at late times during black hole evaporation, a region known as the island emerges inside the black hole and is connected to the Hawking radiation outside. This island contributes to the computation of the entanglement entropy of Hawking radiation, making the entropy consistent with the Page curve and restoring unitarity and information conservation. This approach has been successfully studied and applied to various black hole and gravitational models, such as Schwarzschild black holes \cite{19}, Reissner–Nordström black holes \cite{20}, charged linear dilaton black holes \cite{21}, and noncommutative black holes \cite{22}. \\

The motivation of this work is to investigate the fundamental role of the central charge of the boundary theory in determining the gravitational nature of the bulk, from classical to quantum gravity, besides its consequences for black hole thermodynamics and the information paradox. We aim to establish a clear connection between variations in the central charge and the resulting physical phenomena in the bulk, through both classical and quantum black holes within the AdS/CFT correspondence. \\

We organize this paper as follows. In Section \ref{S2}, we present the holographic description of AdS black holes and discuss the role of the central charge in the boundary CFT. In Section \ref{S3}, we analyze the thermodynamic properties of AdS Schwarzschild black holes in classical and quantum gravity. Section \ref{S4} explores the entanglement entropy of Hawking radiation and the impact of the central charge on the information loss paradox through the island formula. Finally, Section \ref{S5} provides the discussions and conclusions of our results.

\section{Holographic Description}
\label{S2}
\hspace{15pt} In this section, we review the description of AdS black holes within the context of the AdS/CFT correspondence. We also discuss how the variation of the degrees of freedom in the boundary theory changes the nature of gravity in the bulk, and its implications for black hole solutions in the bulk.

\subsection{Holographic Interpretation}
\hspace{15pt} Black holes are a cornerstone of theoretical physics and play a central role in addressing grand questions such as the nature of spacetime  \cite{YY3}, the information loss paradox \cite{1}, the quantum theory of gravity, and the unification of forces  \cite{YY2}. In particular, within the AdS/CFT correspondence, black holes provide an excellent system and laboratory for testing and extracting new theoretical insights. In this context, a gauge theory at the boundary   is related to the gravitational theory in $(d+1)$ AdS spacetime  dimensions by holographic duality.

In this work we focus on the implications of the degrees of freedom of the gauge theory  at the boundary for the bulk, where the partition function of CFT, $Z_{CFT}$, is equivalent to the partition function of AdS, $Z_{\mathrm{AdS}}$ \cite{27}. For a gauge theory at finite temperature, the free energy is related to the partition function as 
\begin{equation}
	W=\mu\, C = -T \ln Z_{\mathrm{CFT}},
	\label{A1}
\end{equation}
where $C$ is the central charge of CFT and is related to the number of degrees of freedom $N$ of CFT by
\begin{equation}
	C \sim N^{2},
	\label{N}
\end{equation}
and $\mu$ represents the chemical potential. On the other hand, the partition function of the bulk is given by the Euclidean path integral $I_E$ \cite{28}
\begin{equation}
	I_E = - \ln Z_{\mathrm{AdS}}.
	\label{A2}
\end{equation}
Combining Eqs.~\eqref{A1} and \eqref{A2} yields the relation between bulk and boundary quantities,
\begin{equation}
	\mu\, C = T\, I_E.
	\label{A3}
\end{equation}

From Eq.~\eqref{A3} we see that the central charge is a fundamental parameter not only for the boundary theory but it also has an extension into bulk quantities. The central charge can be written in terms of bulk parameters \cite{5} as 
\begin{equation}
	C = \frac{\ell^{d-1}}{G},
	\label{C}
\end{equation}
where $\ell$ is the AdS radius, $G$ is Newton's constant, and $d$ is the number of boundary spacetime dimensions. Thus, a variation of the central charge of the boundary gauge theory corresponds to a variation of the AdS radius, or of Newton's constant, or of both. In this work we consider the AdS radius fixed,  i.e. the Newton constant varies \cite{6}. Within this framework the relationship between boundary quantities and those of black hole  in the bulk can be expressed by the following dictionary
\begin{equation}
	\label{HQ}
	\tilde{S}=S,\qquad \tilde{T}=T,\qquad \tilde{E}=M,\qquad
	\tilde{\Phi}= \frac{\Phi \sqrt{G}}{\ell},\qquad
	\tilde{Q}= \frac{Q \ell}{\sqrt{G}},
\end{equation}
where  tilde (without a tilde) refers to boundary quantities (bulk). The symbols \(S\), \(T\), \(\Phi\), \(Q\), \(M\) and \(\tilde{E}\) denote the black hole entropy, Hawking temperature, electric potential, electric charge, black hole mass, and the internal energy of CFT, respectively. The  first law of holographic thermodynamics then takes the form
\begin{equation}
	\label{fl}
	dM = T\, dS + \Phi\, d\tilde{Q} + \mu\, dC.
\end{equation}

\subsection{Central Charge of  Boundary Theories  }
\hspace{15pt} The central charge of CFT is related to the degrees of freedom through Eq.~\eqref{N}, and its variation corresponds to the variation of  the boundary theory. Through the holographic duality between the bulk and the boundary, the central charge of CFT has direct implications for the bulk: it is connected to the Euclidean path integral of the gravitational theory in the bulk and is also related to Newton’s constant and the AdS radius \cite{29}. Furthermore, the value of the central charge  strongly affects the gravitational theory in the bulk.
\\

For large $C$ at the boundary, the bulk corresponds to a classical theory of gravity. Indeed, from Eq.~\eqref{C}, large $C$ implies $G \to 0$, making quantum corrections small and negligible, so that the bulk description is well-approximated by classical general relativity \cite{9,10}. Conversely, for small $C$ at the boundary, the bulk corresponds to a quantum theory of gravity, where quantum corrections are strong and cannot be neglected \cite{11,12,13,14}, pushing the bulk beyond the classical gravity.
On the other hand, in the context of the holographic thermodynamics of black holes, the central charge plays the role of a thermodynamic variable. The value of the central charge at the boundary affects the thermodynamic behavior and phase structure of AdS black holes in the bulk  \cite{A1,A3}. However, the issue with this study lies in the following: for all values of the central charge, the gravitational theory in the bulk is described by classical gravity, and the corresponding black holes are classical black holes \cite{6,7,8,A1,A2,A3,A4,y1,y2,y3,y4,y5}. Nevertheless, these classical black hole solutions are physically meaningful only for large $C$, where quantum corrections in the bulk are negligible \cite{4,zz2}.  For small  or finite $C$, where quantum corrections become significant \cite{qq,qqq}, the classical black hole solutions are no longer valid.
\\


\section{Quantum and Classical Black Holes}
\label{S3}
\hspace{15pt} In this section, we study quantum and classical black holes in the bulk, focusing on the central charge of the boundary CFT. Small central charge corresponds to quantum black holes, while large central charge corresponds to classical one. Quantum black holes are described by Schwarzschild-AdS black holes in loop quantum gravity (LQG), whereas classical black holes are taken to be Schwarzschild-AdS black holes in general relativity. 

\subsection{AdS-Schwarzschild Black Holes in LQG}
\hspace{15pt} The AdS-Schwarzschild black hole solution in loop quantum gravity  is given by \cite{31}
\begin{equation}
	\label{e8}
	ds^2 = - f(r)\, dt^2 + \frac{dr^2}{f(r)} + r^2 d\theta^2 + r^2 \sin^2\theta\, d\phi^2,
\end{equation}
with
\begin{equation}
		\label{e88}
	f(r) = 1 - \frac{2 G M}{r} + \frac{\alpha G^2 M^2}{r^4} + \frac{r^2}{\ell^2},
\end{equation}
where $\ell$ is the AdS radius, and $\alpha = 16 \sqrt{3} \pi \gamma^3 G$ represents the quantum gravity correction in LQG, with $\gamma$ being the Barbero–Immirzi parameter.
\\

The mass of the black hole is obtained by solving $f(r_+) = 0$, where $r_+$ is the event horizon
\begin{equation}
	M_\pm = \frac{r_+^2 \left(\ell r_+ \pm \sqrt{-\alpha \ell^2 - \alpha r_+^2 + \ell^2 r_+^2}\right)}{G \alpha \ell}.
	\label{m}
\end{equation}
\\

There are two branches for the mass, but the positive branch diverges as $\alpha \to 0$, and is therefore discarded. We choose the negative branch. For the mass to be real and positive, $\alpha$ must satisfy
\begin{equation}
	0 < \alpha \leq \frac{\ell^2 r_+^2}{\ell^2 + r_+^2}.
\end{equation}
\\

According the surface gravity, $\kappa$, the Hawking temperature  is given by
\begin{equation}
T= \frac{\kappa}{2 \pi}=\frac{f'(r)}{4 \pi} = \frac{\alpha \left(2 \ell^2 + 3 r_+^2\right) + 3 \ell r_+ \left(\sqrt{\ell^2 r_+^2 - \alpha \left(\ell^2 + r_+^2\right)} - \ell r_+\right)}{2 \pi \alpha \ell^2 r_+}.
\end{equation}
\\

	The entropy of black holes in LQG is given by Rovelli’s approach \cite{32} as  
\begin{equation}
 	S=\frac{b A}{8\, \pi G},
\end{equation}
where $b$ is a constant in the range $0.48 < b < 0.69$,  and $A$ is the area of black hole.
\\

We notice that this entropy follows the Bekenstein-Hawking area law, as it is proportional to the black hole area. However, the entropy of black holes in LQG seems  smaller than the Bekenstein-Hawking entropy in classical gravity $(S = A / 4G)$ due to quantum corrections. This indicates that quantum gravitational effects limit the number of  microstates.

\subsection{Holographic Thermodynamics}
\hspace{15pt} We investigate the thermodynamic quantities of black holes in the context of the AdS/CFT correspondence for small central charge of the boundary theory, where the bulk gravity is described by quantum gravity (LQG). In order to calculate the thermodynamic quantities, we rewrite the following parameters as
\begin{equation}
	\alpha = \sigma G, \qquad r_+ = \sqrt{\frac{2 G S}{b}}, \qquad G = \frac{\ell^2}{C},
\end{equation}
where $\sigma = 16 \sqrt{3} \pi \gamma^3 = 1.1663$, with $\gamma = 0.2375$ \cite{33}. Using these parameters, the mass of the black hole can be expressed as
\begin{equation}
	M = \frac{2 \left(\sqrt{2 C} S^{3/2} - S \sqrt{-\sigma C b - 2 \sigma S + 2 C S}\right)}{\sigma b^{3/2} \ell}.
\end{equation}
\\

Applying the first law of holographic thermodynamics, Eq.~\eqref{fl}, we obtain
\begin{equation}
	T = \left(\frac{\partial M}{\partial S}\right)_C = \frac{2 b \sigma C + 6 \sigma S - 6 C S + 3 \sqrt{2} \sqrt{C S (-b \sigma C - 2 \sigma S + 2 C S)}}{b^{3/2} \sigma \ell \sqrt{-b \sigma C + 2 (-\sigma + C) S}},
\end{equation}
\begin{equation}
	\mu = \left(\frac{\partial M}{\partial C}\right)_S = \frac{2 \left(\frac{S^{3/2}}{\sqrt{2} \sqrt{C}} + \frac{(b \sigma - 2 S) S}{2 \sqrt{-b \sigma C + 2 (-\sigma + C) S}}\right)}{b^{3/2} \sigma \ell}.
\end{equation}

To study the thermodynamic behavior of black holes in LQG for small boundary degrees of freedom, and to investigate the impact of the nature of gravity on emergent bulk phenomena, we analyze the thermal evolution and heat capacity.  Firstly, we determine the critical points by solving
\begin{equation}
	\left( \frac{\partial T}{\partial S} \right)_C = 0, \qquad
	\left( \frac{\partial^2 T}{\partial S^2} \right)_C = 0.
\end{equation}
The critical quantities are
\begin{equation}
	S_c = \frac{512 b \sigma}{243}, \qquad C_c = \frac{256 \sigma}{13}.
\end{equation}

We observe that there is a critical point in the thermal evolution of Schwarzschild black holes in LQG, which does not exist for classical Schwarzschild–AdS black holes. However, in LQG, black holes correspond only to small values of the central charge $(C \to 0)$, so the central charge is always smaller than the critical value. Therefore, the critical phenomenon observed in classical gravity does not occur in LQG. We conclude that the absence of critical phenomenon in Schwarzschild black holes is universal, independent of the nature of gravity (classical or quantum) in the bulk, and instead depends on the physical properties of the black hole (electric charge, rotation, etc.).

\bigskip

To study the stability and the phase transition, we use the heat capacity expression given by 
\begin{equation}
	\zeta = T \left( \frac{\partial S}{\partial T}\right)_C = \frac{
		128 \, b \, \sigma \, \sqrt{c} \, \sqrt{s} \, \big(c (1024 s - 243) - 52 s \big)^{3/2} \Psi
	}{
		243 \left( \Lambda - 2 c \big(-c (1024 s - 243) + 52 s \big)^2 \right)
	}.
\end{equation}
where 
\begin{equation}
	\Psi = \left( -16 \sqrt{c} \sqrt{s} \sqrt{1024 c s - 243 c - 52 s} + c (512 s - 81) - 26 s \right),
\end{equation}
\begin{equation}
	\Lambda = \sqrt{c} \sqrt{s} (256 c - 13) \big(c (256 s - 81) - 13 s\big) \sqrt{c (1024 s - 243) - 52 s},
\end{equation}
and 
\begin{equation}
	s = \frac{S}{S_c}, \qquad c = \frac{C}{C_c}.
\end{equation}

\begin{figure}
	\centering
	\includegraphics[width=0.7\linewidth]{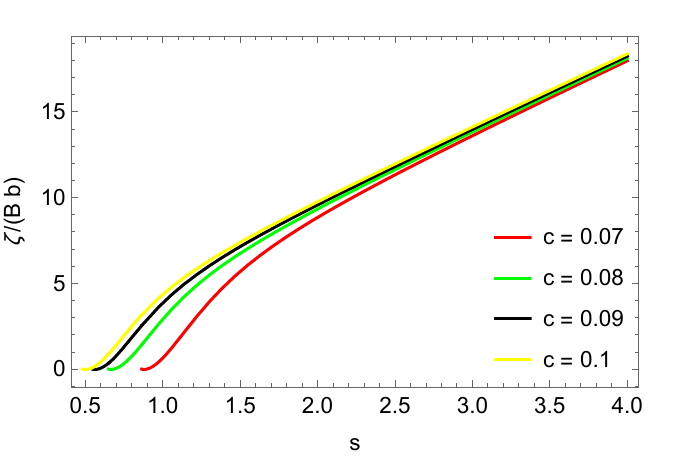}
	\caption{Heat capacity normalized as a function of entropy for various values of the central charge.}
	\label{hc}
\end{figure}

Figure~\ref{hc} shows the evolution of the heat capacity of AdS–Schwarzschild black holes in loop quantum gravity as a function of entropy for different small values of the central charge of CFT. We observe that the heat capacity is always positive for $b > 1$ and $\sigma > 1$, which indicates that  black holes are stable and that no phase transition occurs.

\subsection{Classical Black Holes}
\hspace{15pt} The transition from quantum to classical gravity is a major challenge in theoretical physics and is not yet fully understood. In this paper, the quantum black hole (Schwarzschild black hole in LQG) metric is given by Eqs.~\eqref{e8} and \eqref{e88}, while the classical black hole (AdS–Schwarzschild black hole in GR) metric function is written as
\begin{equation}
	\label{cm}
	f(r) = 1 - \frac{2 G M}{r} + \frac{r^2}{\ell^2}.
\end{equation}

We can obtain the classical black hole solution from the quantum solution by taking the appropriate limit: large $C$ (small $G$) on the boundary corresponds to classical gravity in the bulk, while small $C$ (large $G$) corresponds to quantum gravity in the bulk. For sufficiently small $G$ (so that $G^3 \ll G$), the quantum correction term proportional to $G^3$ can be neglected in the quantum black hole solution, and one recovers the classical black hole solution.

\subsubsection{Classical Black Hole Thermodynamics}
\hspace{15pt} We investigate the thermodynamics of classical black holes in the context of holographic thermodynamics to compare the classical and quantum black holes. The mass of classical black holes in terms of holographic quantities can be written by solving $f(r_+) = 0$ from Eq.~\eqref{cm} 
\begin{equation}
	M = \frac{S^{2} + \pi S C}{2 \pi^{3/2} \ell (S C)^{1/2}}.	
\end{equation}
\\

Through the first law, Eq.~\eqref{fl}, we obtain the other thermodynamic quantities as 
\begin{equation}
	T = \left(\frac{\partial M}{\partial S}\right)_C = \frac{3S^{2} + \pi S C}{4 \pi^{3/2} \ell S (S C)^{1/2}},
\end{equation}
\begin{equation}
	\mu = \left(\frac{\partial M}{\partial C}\right)_S = -\frac{S^{2} - \pi S C}{4 \pi^{3/2} \ell C (S C)^{1/2}}.
\end{equation}
\\

The heat capacity is given by
\begin{equation}
	\zeta = T \left( \frac{\partial S}{\partial T}\right)_C = \frac{2S(3S + \pi C)}{3S - \pi C}.
	\label{hcz}
\end{equation}

From the expression of the heat capacity, Eq. \eqref{hcz}, we observe that small black holes ($S < \pi C/3$) are unstable, as they have negative heat capacity, while large black holes  are stable, having positive heat capacity. Thus, the thermodynamic behavior of AdS–Schwarzschild black holes in classical gravity differs from that of AdS–Schwarzschild black holes in LQG, particularly regarding stability and phase transition. Quantum black holes are always stable, whereas classical black holes are unstable for small sizes and stable for large ones. We conclude that the nature  gravity in the bulk affects the stability of black holes.

\section{Entanglement Entropy and Central Charge}
 \label{S4}
\hspace{15pt} In this section, we focus on the role of the central charge of the boundary theory within the framework of the information loss paradox. The process of Hawking radiation emission appears to conflict with quantum mechanics, as black hole evaporation seems to violate unitarity. Page proposed that the entanglement entropy of Hawking radiation must follow the so-called Page curve in order to preserve unitarity. This curve is characterized by a specific Page time, before and after which the entanglement entropy of Hawking radiation increases and decreases, respectively.
\\

Recently, a solution to this paradox has been proposed through consideration of an emergence of a region inside the black hole, called the \emph{island}\footnote{The island is a region inside the black hole that emerges from the replica wormhole saddle points in the gravitational path integral, connecting the Hawking radiation outside the black hole with a part of its interior.}. This region is taken into account when calculating the entanglement entropy of Hawking radiation by means of  the island formula  expressed as \cite{18}
\begin{equation}
	S_{gen} (R)=\min\left\{\operatorname{ext}\!\left[\frac{\mathrm{Area}(\partial I)}{4G}+S_{\mathrm{Bulk}}(R\cup I)\right]\right\}, 
\end{equation}
where $R$ is the region of radiation outside the black hole, $I$ is the island, and $\partial I$ its  boundary.  
	Furthermore, ``ext'' denotes the extremization of the generalized entropy with respect to the position of the island boundary $\partial I$, i.e. solving 
	$\partial S_{\mathrm{gen}} / \partial x_i = 0$ (see below Eqs.~\eqref{EE1} and \eqref{EE2}). 
	The notation ``min'' indicates that among all extremal solutions, we choose the one corresponding to the minimal entropy.

We start by studying the entanglement entropy without the inclusion of islands, for both classical and quantum black holes.

To simplify the calculation, Kruskal coordinates are useful \cite{YA}
\begin{equation}
	U = -e^{-\kappa (t - r_*)}, \qquad V = e^{\kappa (t + r_*)},
\end{equation}
where $\kappa$ is the surface gravity, and $r_*$ is the tortoise coordinate given by 
\begin{equation}
	r_* = \int^{r}_{0} \frac{1}{f(r)} \, dr.
\end{equation}

The metric of black holes can be expressed in terms of these coordinates as follows \cite{YA}
\begin{equation}
	ds^2 = -\Omega^2(r) \, dU  dV,
\end{equation} 
where $\Omega(r)$ is the conformal factor 
 expressed as
\begin{equation}
	\Omega^2(r) = \frac{f(r)}{\kappa^2 e^{2 \kappa r_*}}.
\end{equation}
\\

To describe the Hawking radiation as observed by a distant observer, one can use the s-wave approximation\footnote{{The s-wave approximation allows replacing the 4d CFT with a 2d CFT to describe Hawking radiation at large distances, neglecting the angular components of the metric, where the spacetime at large distances becomes  flat.}} to neglect the spherical part of the metric.
\\

\begin{figure}[htbp!]
	\centering
	\subfloat[Quantum Black Hole]{%
		\includegraphics[width=0.45\linewidth]{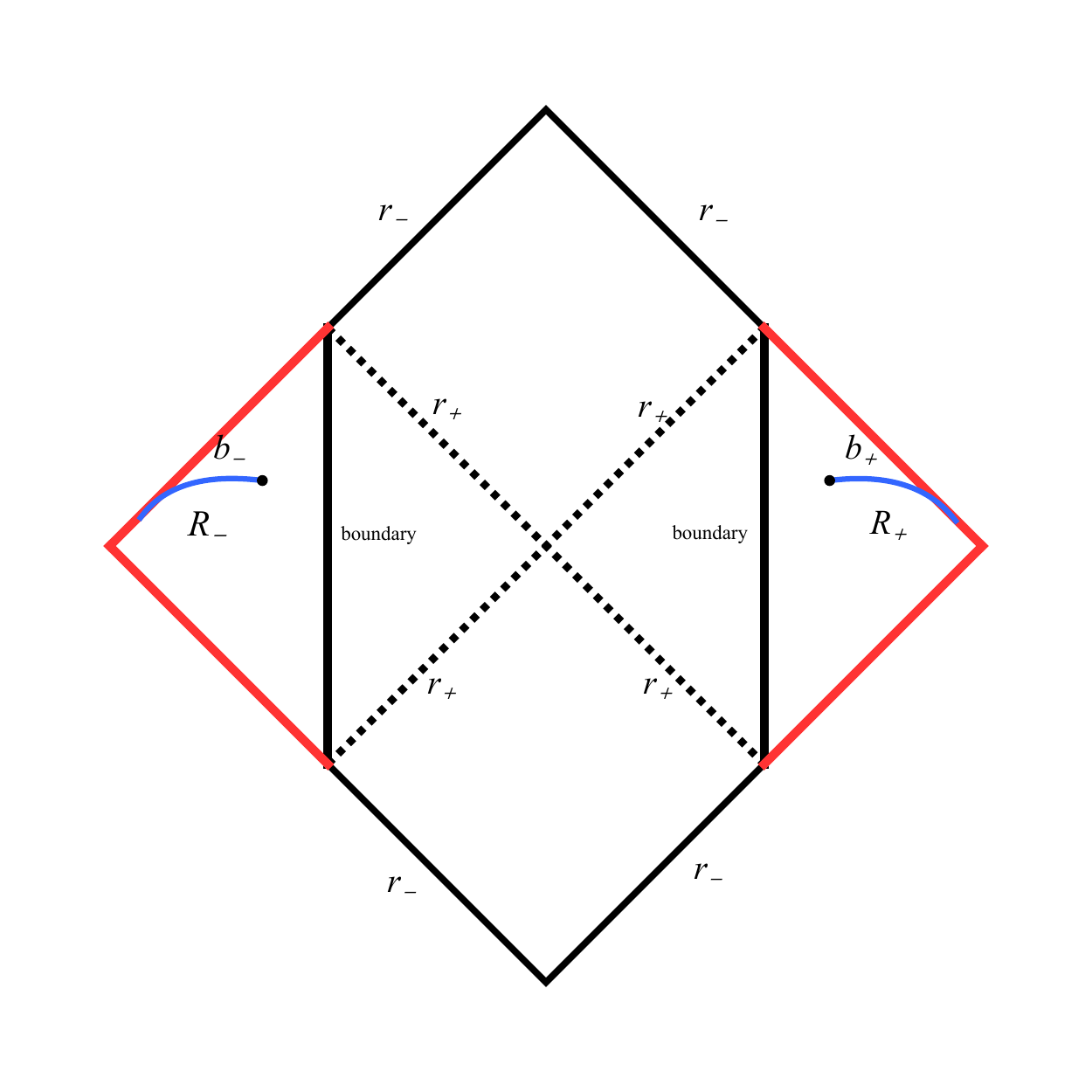}%
		\label{QP}%
	}\hfill
	\subfloat[Classical Black Hole]{%
		\includegraphics[width=0.5\linewidth]{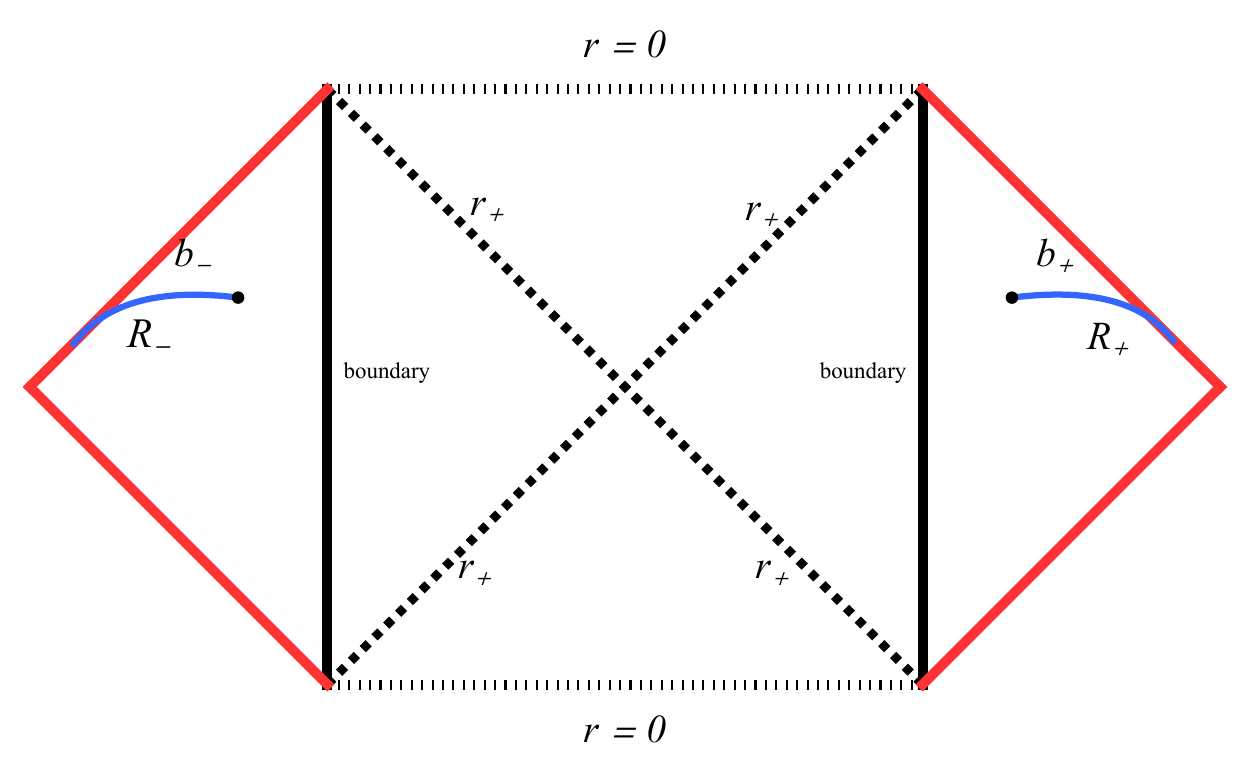}%
		\label{CP}%
	}
	\caption{Penrose diagrams of classical and quantum black holes without island.}
	\label{PD}
\end{figure}

Figure~\ref{PD} shows the Penrose diagrams of classical and quantum black holes without  the island. Here,  $R_\pm$ represent the regions of Hawking radiation in the right and left wedges, and $b_\pm$ denote the boundaries of the regions $R_\pm$. The coordinates of $b_+$ are $ (t_b, b)$ and those of $b_-$ are $(t, r) = (-t_b + i \beta/2, b)$, where $\beta$ is the inverse temperature expressed in terms of the surface gravity, $\beta = 2\pi/\kappa$. We assume that the entire system is in a pure state at $t = 0$. 
\\

The entanglement entropy of Hawking radiation is given by \cite{34}
\begin{equation}
	S(R)= \frac{C}{3} \log d(b_+,b_-),
\end{equation} 
where $d(b_+,b_-)$ is the geodesic distance, expressed as \cite{YA}
\begin{equation}
	\label{gd}
	d^2 (b_+,b_-) = \Omega^2(b) \, [U(b_-) - U(b_+)] [V(b_+) - V(b_-)].
\end{equation} 
At large distances, $f(b) = 1$. Thus, the entanglement entropy of Hawking radiation  reads as
\begin{equation}
	S(R)= \frac{C}{3} \log \left( \frac{2 \cosh \left( \kappa\, t_b \right) }{\kappa}\right).
\end{equation}  

For $t_b \to \infty$, the entanglement entropy of Hawking radiation becomes
\begin{equation}
	\label{ap}
	S(R)= \frac{C \kappa t_b}{3}.
\end{equation}
\\

From the final expression of the entanglement entropy of Hawking radiation without the island, we see that at late times this entropy diverges and violates the Page curve. Therefore, the state of the black hole at late times is mixed or thermal, i.e. unitarity of black hole evaporation is violated, implying a loss of information. This result holds for both quantum and classical black holes.
\\

\begin{figure}[h!]
	\centering
	\subfloat[Quantum Black Hole]{%
		\includegraphics[width=0.45\linewidth]{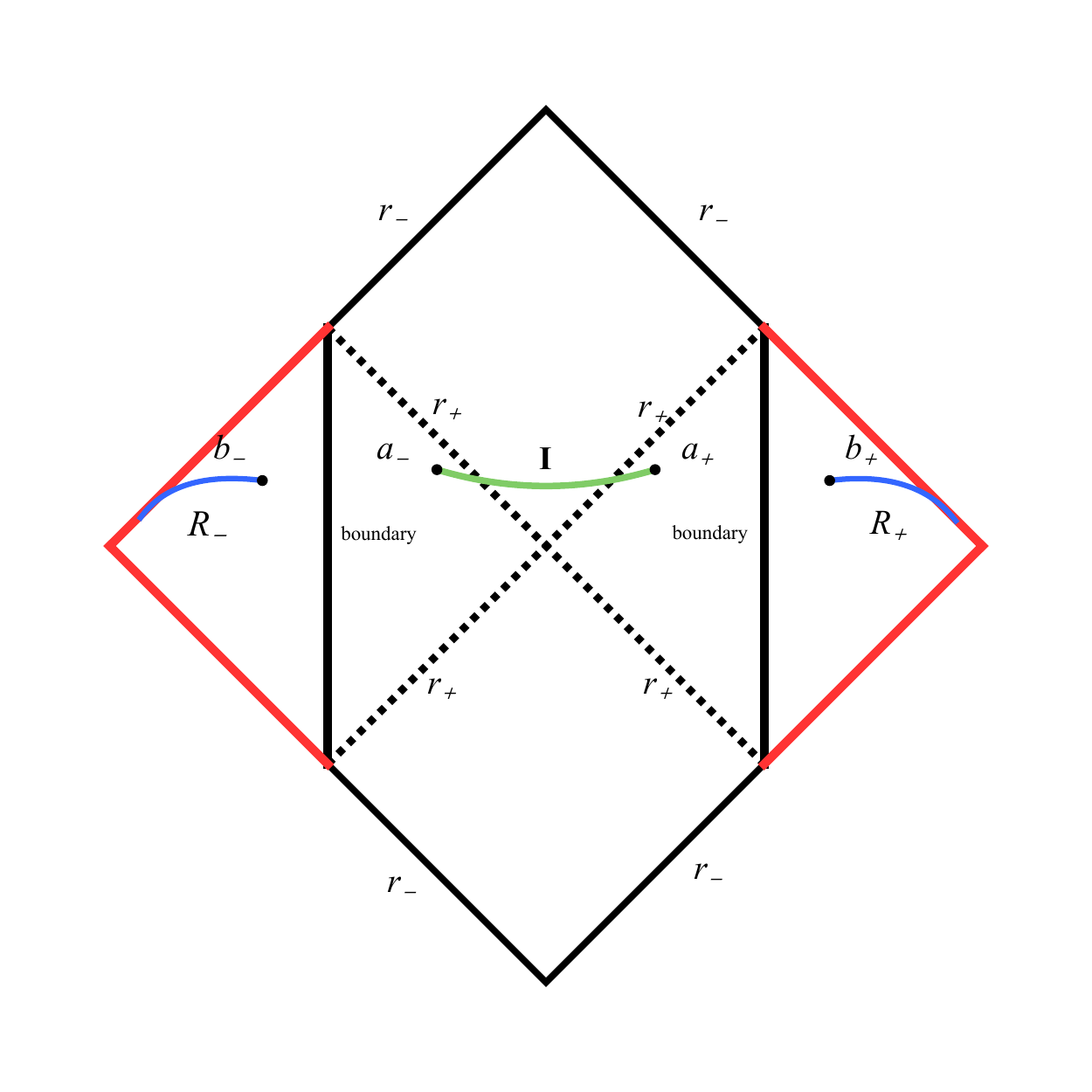}%
		\label{QPI}%
	}\hfill
	\subfloat[Classical Black Hole]{%
		\includegraphics[width=0.5\linewidth]{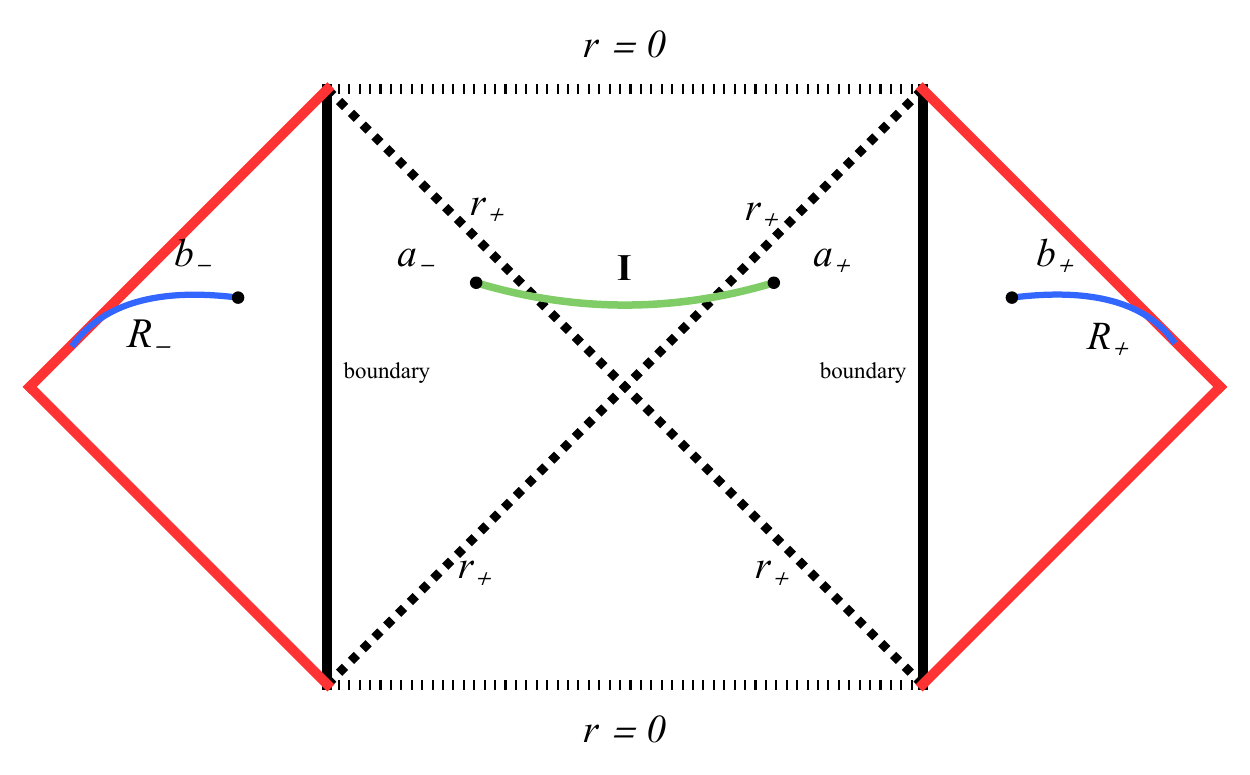}%
		\label{CPI}%
	}
	\caption{Penrose diagrams of classical and quantum black holes with island, represented by the green curve.}
	\label{PDI}
\end{figure}

After the Page time, we include the contribution of the island in the generalized entropy of Hawking radiation, and it is expressed as~\cite{21}.

\begin{equation}
	S_{\mathrm{gen}}(R) = \frac{2\pi a^2}{G} 
	+ \frac{C}{3} 
	\log \frac{
		d(a_+, a_-) \, d(b_+, b_-) \, d(a_+, b_+) \, d(a_-, b_-)
	}{
		d(a_+, b_-) \, d(a_-, b_+)
	},
\end{equation}
where the island boundaries are located at
$a_+ = (t_a, a)$ and $a_- = (-t_a + i\beta / 2, a)$,and $d(a_\pm, b_\pm)$ represents the correlations between the island and the Hawking radiation outside the black hole.
This entropy can be simplified using the late-time approximation, where the distance between the right and left wedges becomes very large~\cite{21}. In this limit, the geodesic distances between events in the left and right wedges become approximately equal, leading to the following relationships
\begin{equation}
	d(a_{+}, a_{-}) \simeq d(b_{+}, b_{-}) \simeq d(a_{+}, b_{-})
	\simeq d(a_{-}, b_{+}) \gg d(a_{+}, b_{+}) \simeq d(a_{-}, b_{-}) .
\end{equation}

Furthermore, there are no correlations between these events, meaning that they do not contribute to the generalized entropy. Using this approximation, we obtain

\begin{equation}
	\label{APPP}
	S_{\mathrm{gen}}(R) = \frac{2\pi a^2}{G} + \frac{C}{3} 
	\log \left[ d(a_+, b_+) \, d(a_-, b_-) \right] ,
\end{equation}
where the geodesics $d(a_\pm, b_\pm)$ are given through Eq.~\eqref{gd} as follows ( see Also Appendix~\ref{AP1})
\begin{equation}
	d(a_+, b_+) = d(a_-, b_-) = 
	\sqrt{\frac{2 \sqrt{f(a)} \Big(\cosh\!\big(\kappa (r_*(a) - r_*(b))\big) - \cosh\!\big(\kappa (t_a - t_b)\big)\Big)}{\kappa^2}}.
\end{equation}
\\

Thus, the generalized entropy is expressed as
\begin{equation}
	\label{46}
	S_{\mathrm{gen}}(R) = \frac{2\pi a^2}{G} 
	+ \frac{C}{3} 
	\log \left( 
	\frac{2\sqrt{f(a)}}{\kappa^2}
	\Big[ 
	\cosh\!\big(\kappa (r_*(a) - r_*(b))\big) 
	- \cosh\!\big(\kappa (t_a - t_b)\big)
	\Big]
	\right).
\end{equation}

To determine the position of the island boundary coordinates $(t_a, a)$ with respect to the extremization of the entanglement entropy of Hawking radiation, we solve 
$\partial S_{\mathrm{gen}} / \partial x_i = 0$, where $x_i = \{t_a, a\}$. 
The extremization with respect to the temporal coordinate reads
\begin{equation}
	\label{EE1}
	\dfrac{\partial S_{\mathrm{gen}}(R)}{\partial t_a}
	=-\frac{C \, \kappa \sinh\left(\kappa(t_a - t_b)\right)}{3 \left(\cosh\left(\kappa(r_*(a) - r_*(b))\right) - \cosh\left(\kappa(t_a - t_b)\right)\right)}=0,
\end{equation}
which gives $t_a = t_b$ as the extremal time of the island boundary. 
Using this result, the extremization with respect to the spatial coordinate is given by (see also Appendix~\ref{AP3})
\begin{equation}
	\label{EE2}
	\dfrac{\partial S_{\mathrm{gen}}(R)}{\partial a}= \frac{4 \pi a}{G} + \frac{C}{6} \frac{f'(a)}{f(a)}+ \frac{C \kappa}{3} \coth\left( \frac{\kappa}{2} \left( r_*(a) - r_*(b) \right) \right) r'_*(a)=0,
\end{equation}
which implies
\begin{equation}
	\label{CG}
	C G= - \frac{24 \pi a f(a)}{2 \kappa f(a)\, r'_*(a) \coth\left( \frac{\kappa}{2}  \left( r_*(a) - r_*(b) \right) \right) + f'(a)}.
\end{equation}

Since the location of the island is near the event horizon of the black hole, we can rewrite 
the spatial coordinate as
\begin{equation}
	\label{APP}
	a \approx r_+ + \epsilon^2 r_+, 
\end{equation}
where we have chosen the squared parameter $\epsilon$ for convenience ($\epsilon^2 \ll 1$). 
We now determine the expression for the correction term $\epsilon$ using the following setup.
\\

Based on the approximation in Eq.~\eqref{APP}, the metric function $f(a)$ can be simplified as
\begin{equation}
	\label{AP}
	f(a) \approx f'(r_+)(a - r_+) = 2 \kappa r_+ \epsilon^2,
\end{equation}
where  the surface gravity $\kappa = f'(r_+)/2$.
Using the approximation in Eq.~\eqref{AP}, we obtain the tortoise coordinate in terms of the surface gravity and the parameter $\epsilon$ as follows  
\begin{equation}
\label{TTz}
	r_*(a) = \int_{0}^{a} \frac{dr}{f(r)} \approx  \frac{1}{\kappa} \log \epsilon.
\end{equation}

By substituting the expressions obtained in Eqs.~\eqref{APP}–\eqref{TTz} into Eq.~\eqref{CG}, we obtain

\begin{equation}
	\label{APA}
	C G = \frac{-12 \pi \epsilon r_+^2 \left( \epsilon e^{-\kappa r_*(b)} - 1 \right)}{e^{-\kappa r_*(b)}} .
\end{equation}

To a first order in \(\epsilon\), the expression for the correction term \(\epsilon\), associated with the spatial coordinate of the island, reads

\begin{equation}
	\label{47}
	\epsilon \approx  \frac{C G  e^{- \kappa r_*(b)}}{ 12 \pi r_+^2}.
\end{equation}

 Finally, we can express the entanglement entropy of Hawking radiation after the Page time using Eq.~\eqref{APP} as 

\begin{equation}
	\label{app}
	S_{\mathrm{gen}}(R)= 2 S + 4 \epsilon^2 S +\frac{C}{6} \log\left(\frac{2  r_+ \left( e^{2 k r_*(b)} - 4  \epsilon e^{k r_*(b)} + 6 \epsilon^2
		\right) }{\kappa^3} \right),
\end{equation} 
where \(S\) is the Bekenstein-Hawking entropy. We conclude that the entanglement entropy of Hawking radiation, considering the island, is time-independent and constant. Furthermore, the entanglement entropy  respects the Page curve, so the black hole evaporation process is unitary, ensuring information conservation in black holes.
\\

For the role of the boundary central charge on the information paradox, especially the Hawking radiation entanglement entropy, we have within the island formula two entropy expressions before and after the Page time. Before the Page time, there is only the contribution of the entanglement entropy of Hawking radiation outside the black hole. The entropy at this stage is given by Eq.~\eqref{ap}, where this entropy is related to three parameters: the measurement time of the entropy (which depends on the observer), the surface gravity (related to the Hawking temperature and the geometry of the black hole), and the central charge (related to the boundary theory and the nature of gravity in the bulk). The central charge in the entanglement entropy before the Page time represents the slope. For small central charge, corresponding to quantum gravity, the evolution of the entanglement entropy is slow. 
\\

For large central charge, corresponding to classical gravity, the entanglement entropy evolves rapidly, and the quantum information spreads quickly. This demonstrates that the central charge significantly impacts the entanglement entropy of Hawking radiation, showing that the evolution of quantum information is influenced by the nature of gravity and the spacetime structure in the bulk. 
\\

After the Page time, i.e. after taking into account the contribution of the island in the entanglement entropy of Hawking radiation (Eq.~\eqref{app}), we find that this entropy is conserved and related to the Bekenstein–Hawking entropy of the black hole and the central charge through the logarithmic term. For small central charge, the logarithmic term is negligible and  the entropy becomes
\begin{equation}
	S_{\mathrm{gen}}(R) \approx 2 S,
\end{equation}
whereas for large central charge, the logarithmic term cannot be neglected, and the entropy is given by
\begin{equation}
	S_{\mathrm{gen}}(R) \approx 2 S + \frac{C}{6} \log \left( \frac{2 e^{2 \kappa r_*(b)} r_+}{\kappa^3} \right).
\end{equation} 
\section{Discussion and Conclusion}
 \label{S5}
\hspace{15pt} In this paper, we investigated, in the context of the AdS/CFT correspondence, the relationship between the degrees of freedom of the boundary theory, represented by the central charge, and the nature of gravity and spacetime in the bulk. The central charge plays a fundamental role in determining whether gravity in the bulk behaves classically or quantum mechanically. Furthermore, it has significant implications for black hole thermodynamics and the  information loss paradox. \\

For small central charge at the boundary, where the bulk spacetime is described by quantum gravity, we use Loop Quantum Gravity (LQG) as the quantum gravity framework. In this case, quantum corrections in the bulk become significant, leading to modifications in the black hole solutions and their thermodynamic behavior. We find that the entropy of AdS-Schwarzschild black holes in LQG satisfies the Bekenstein-Hawking area law, being proportional to the event horizon area of the black hole, but smaller than the standard Bekenstein-Hawking entropy due to quantum corrections. Furthermore, regarding the thermodynamic behavior of quantum AdS-Schwarzschild black holes, the heat capacity is always positive indicating that black holes are always stable, with no phase transition occurring. \\

Conversely, for large central charge of CFT, where the bulk is described by classical gravity, i.e. quantum corrections are negligible, the AdS–Schwarzschild black holes exhibit two thermodynamic phases: stable and unstable. The large black holes are stable and have positive heat capacity, while small black holes are unstable with negative heat capacity. \\

We also explored the relationship between the central charge and the information loss paradox, particularly the impact of the central charge on the entanglement entropy of Hawking radiation. This entropy always increases with time, violating unitarity and not following the Page curve. To address this issue, we use a recent approach, the ``island formula,'' in which, before the Page time, the entanglement entropy of Hawking radiation is determined solely by  radiation outside the black hole, while after the Page time, the island is taken into account in the entanglement entropy. Before the Page time, the entropy increases linearly with time, and the central charge determines the slope of this increase. For small central charge, corresponding to quantum gravity in the bulk, the entanglement entropy increases slowly, implying that quantum information spreads slowly. For large central charge, corresponding to classical gravity, the entanglement entropy of Hawking radiation increases rapidly.
 \\


After the Page time, when the island emerges and is included in the calculation of the entanglement entropy of Hawking radiation, the entropy becomes constant, respecting unitarity and the conservation of information. Moreover, the resulting entropy depends on the Bekenstein–Hawking entropy of the black hole and a logarithmic correction related to the central charge of CFT. For small central charge, the logarithmic correction can be neglected, and the entropy becomes approximately twice the Bekenstein–Hawking entropy. However, for large central charge, the logarithmic term cannot be neglected. \\

In summary, our analysis reveals that the central charge of the boundary theory acts as a key holographic parameter governing the nature of gravity and the bulk geometry within the AdS/CFT framework, as well as the thermodynamic behavior and the entanglement entropy of Hawking radiation and quantum information. This study also highlights how the properties of the boundary theory influence the AdS/CFT correspondence in the gravitational system and the physical phenomena in the bulk.
\appendix
\numberwithin{equation}{section}
\section{Generalized Entropy}
\label{AP1}

In this appendix, we derive some equations related to the generalized entropy. Based on \cite{YA}, the geodesic distance is expressed as

\begin{equation}
	d^2(a_+,b_+) = \Omega(a)\,\Omega(b)\,
	\left[ U(b_+)- U(a_+) \right]
	\left[ V(a_+) - V(b_+) \right],
\end{equation}

where the conformal factor is given by

\begin{equation}
	\Omega(a) = \frac{\sqrt{f(a)}}{\kappa e^{\kappa r_*(a)}}, \qquad
	\Omega(b) = \frac{1}{\kappa e^{\kappa r_*(b)}},
\end{equation}

and Kruskal coordinates are expressed as follows

\begin{equation}
	U(b_+) = -e^{-\kappa (t_b - r_*(b))}, \qquad
	V(b_+) = e^{\kappa (t_b + r_*(b))},
\end{equation}

\begin{equation}
	U(a_+) = -e^{-\kappa (t_a - r_*(a))}, \qquad
	V(a_+) = e^{\kappa (t_a + r_*(a))}.
\end{equation}

Finally,

\begin{equation}
	\begin{split}
		d^2(a_+,b_+) &= 
		\frac{\sqrt{f(a)}}{\kappa^2 e^{\kappa ( r_*(a)+ r_*(b))}}
		\left[
		-e^{-\kappa (t_b - r_*(b))} + e^{-\kappa (t_a - r_*(a))}
		\right]
		\left[
		e^{\kappa (t_a + r_*(a))} - e^{\kappa (t_b + r_*(b))}
		\right] \\
		&= 
		\frac{2\sqrt{f(a)}}{\kappa^2}
		\Big[
		\cosh\!\big(\kappa (r_*(a) - r_*(b))\big)
		- \cosh\!\big(\kappa (t_a - t_b)\big)
		\Big],
	\end{split}
\end{equation}

For $d(a_-,b_-)$, we have

\begin{equation}
	d^2(a_-,b_-) = \Omega(a)\,\Omega(b)\,
	\left[ U(b_-) - U(a_-) \right]
	\left[ V(a_-) - V(b_-) \right],
\end{equation}

where the Kruskal coordinates are expressed as follows

\begin{equation}
	U(b_-) = -e^{-\kappa (-t_b + i\beta / 2 - r_*(b))}, \qquad
	V(b_-) = e^{\kappa (-t_b + i\beta / 2 + r_*(b))},
\end{equation}

\begin{equation}
	U(a_-) = -e^{-\kappa (-t_a + i\beta / 2 - r_*(a))}, \qquad
	V(a_-) = e^{\kappa (-t_a + i\beta / 2 + r_*(a))}.
\end{equation}

where $\beta = 2\pi/\kappa$. We find

\begin{equation}
	\begin{split}
		d^2(a_-,b_-) &= 
		\frac{\sqrt{f(a)}}{\kappa^2 e^{\kappa ( r_*(a)+ r_*(b))}}
		\left[
		-e^{-\kappa (t_b - r_*(b))} + e^{-\kappa (t_a - r_*(a))}
		\right]
		\left[
		e^{\kappa (t_a + r_*(a))} - e^{\kappa (t_b + r_*(b))}
		\right] \\
		&= 
		\frac{2\sqrt{f(a)}}{\kappa^2}
		\Big[
		\cosh\!\big(\kappa (r_*(a) - r_*(b))\big)
		- \cosh\!\big(\kappa (t_a - t_b)\big)
		\Big].
	\end{split}
\end{equation}

\section{Extremal Condition}
\label{AP3}

The generalized entropy is expressed as

\begin{equation}
	S_{\mathrm{gen}}(R) =
	\frac{2\pi a^2}{G}
	+ \frac{C}{3}
	\log \left(
	\frac{2\sqrt{f(a)}}{\kappa^2}
	\Big[
	\cosh\!\big(\kappa (r_*(a) - r_*(b))\big)
	- \cosh\!\big(\kappa (t_a - t_b)\big)
	\Big]
	\right),
\end{equation}

For $t_a = t_b$, we find

\begin{equation}
	S_{\mathrm{gen}}(R) =
	\frac{2\pi a^2}{G}
	+ \frac{C}{3}
	\log \left(
	\frac{2\sqrt{f(a)}}{\kappa^2}
	\Big[
	\cosh\!\big(\kappa (r_*(a) - r_*(b))\big)
	- 1
	\Big]
	\right),
\end{equation}

Thus, $\dfrac{\partial S_{\mathrm{gen}}(R)}{\partial a}$ is given by

\begin{equation}
	\dfrac{\partial S_{\mathrm{gen}}(R)}{\partial a}
	= \frac{4 \pi a}{G}
	+ \frac{C}{6} \frac{f'(a)}{f(a)}
	+ \frac{C}{3}
	\frac{\kappa
		\sinh \!\left[\kappa \left( r_{*}(a) - r_{*}(b) \right)\right]
		r_{*}'(a)}
	{\cosh \!\left[\kappa \left( r_{*}(a) - r_{*}(b) \right)\right] - 1},
\end{equation}

We know that

\begin{equation}
	\frac{\sinh x}{\cosh x - 1} =
	\coth\!\left(\frac{x}{2}\right),
\end{equation}

Using this identity, we can write

\begin{equation}
	\frac{\sinh\!\left(\kappa\left(r_{*}(a)-r_{*}(b)\right)\right)}
	{\cosh\!\left(\kappa\left(r_{*}(a)-r_{*}(b)\right)\right)-1} =
	\coth\!\left(\frac{\kappa}{2}
	\left(r_{*}(a)-r_{*}(b)\right)\right),
\end{equation}

Finally,

\begin{equation}
	\label{EE2}
	\dfrac{\partial S_{\mathrm{gen}}(R)}{\partial a}
	= \frac{4 \pi a}{G}
	+ \frac{C}{6} \frac{f'(a)}{f(a)}
	+ \frac{C \kappa}{3}
	\coth\!\left(
	\frac{\kappa}{2}
	\left( r_*(a) - r_*(b) \right)
	\right)
	r'_*(a).
\end{equation}

\section*{Acknowledgments}
Y. Ladghami gratefully acknowledges the support from the "PhD-Associate Scholarship – PASS" grant provided by the National Center for Scientific and Technical Research in Morocco, under grant number 42 UMP2023.


\begin{thebibliography}{9}
	\bibitem{1} S. W. Hawking, 
{\em Communications in Mathematical Physics}, \textbf{43}, 199–220 (1975).
\bibitem{2} J. D. Bekenstein, 
{\em Physical Review D}, \textbf{7}, 2333–2346 (1973).
\bibitem{3} J. Bardeen, B. Carter, and S. W. Hawking, 
{\em Communications in Mathematical Physics}, \textbf{31}, 161–170 (1973).
\bibitem{4} J. Maldacena, 
{\em International Journal of Theoretical Physics}, \textbf{38}, 1113–1133 (1999)
\bibitem{5} M. R. Visser, 
{\em Physical Review D}, \textbf{105}, 106014 (2022).
\bibitem{6} Z. Gao and L. Zhao, 
{\em Classical and Quantum Gravity}, \textbf{39}(7), 075019 (2022).
\bibitem{7} Y. Ladghami, B. Asfour, A. Bouali, A. Errahmani, and T. Ouali, 
{\em Physics of the Dark Universe}, \textbf{44}, 101470 (2024).
\bibitem{8} Y. Ladghami, B. Asfour, A. Bouali, A. Errahmani, and T. Ouali, 
{\em Physics of the Dark Universe}, \textbf{41}, 101261 (2023).
\bibitem{A1} Y. Ladghami and T. Ouali, 
{\em Physics of the Dark Universe}, \textbf{44}, 101471 (2024).
\bibitem{A3} Y. Ladghami and T. Ouali, 
{\em Universe}, \textbf{11}(10), 337 (2025).
\bibitem{A2} Y. Ladghami, B. Asfour, A. Bouali, A. Errahmani, and T. Ouali, 
{\em Physics Letters B}, \textbf{864}, 139418 (2025).
\bibitem{A4} Y. Ladghami, B. Asfour, A. Bouali, T. Ouali, and G. Mustafa, 
{\em Physics of the Dark Universe}, \textbf{46}, 101724 (2024).

\bibitem{9} E. Witten, 
{\em Advances in Theoretical and Mathematical Physics}, \textbf{2}(2), 253–290 (1998).
\bibitem{10} S. S. Gubser, I. R. Klebanov, and A. M. Polyakov, 
{\em Physics Letters B}, \textbf{428}(1-2), 105–114 (1998).
\bibitem{11} J. D. Brown and M. Henneaux, 
{\em Communications in Mathematical Physics}, \textbf{104}(2), 207–226 (1986).
\bibitem{12} A. Strominger, 
{\em Journal of High Energy Physics}, \textbf{1998}(02), 009 (1998).
\bibitem{13} S. D. Mathur, 
{\em Fortschritte der Physik: Progress of Physics}, \textbf{53}(7-8), 793–827 (2005).
\bibitem{14} G. T. Horowitz and J. Polchinski, 
{\em Approaches to Quantum Gravity}, 169–186 (2009).
\bibitem{15} C. Rovelli, 
{\em Living Reviews in Relativity}, \textbf{11}(1), 5 (2008).
\bibitem{16} S. Raju, 
{\em Physics Reports}, \textbf{943}, 1–80 (2022).
\bibitem{17} D. N. Page, 
{\em Physical Review Letters}, \textbf{71}(23), 3743 (1993).
\bibitem{AMPS2013}
A.~Almheiri, D.~Marolf, J.~Polchinski, J.~Sully,
{\em Journal of High Energy Physics}, \textbf{02}, 062 (2013).
\bibitem{18} A. Almheiri, T. Hartman, J. Maldacena, E. Shaghoulian, and A. Tajdini, 
{\em Reviews of Modern Physics}, \textbf{93}(3), 035002 (2021).
\bibitem{19} K. Hashimoto, N. Iizuka, and Y. Matsuo, 
{\em Journal of High Energy Physics}, \textbf{2020}(6), 1–21 (2020).
\bibitem{20} Y. Ling, Y. Liu, and Z. Y. Xian, 
{\em Journal of High Energy Physics}, \textbf{2021}(3), 1–23 (2021).
\bibitem{21} B. Ahn, S. E. Bak, H. S. Jeong, K. Y. Kim, and Y. W. Sun, 
{\em Physical Review D}, \textbf{105}(4), 046012 (2022).
\bibitem{22} Y. Liu, W. Xu, and B. Zhang, 
{\em Physics Letters B}, \textbf{139546} (2025).


\bibitem{YY3} J. A. Wheeler and K. Ford,
{\em Geons, Black Holes, and Quantum Foam: A Life in Physics},
W. W. Norton \& Company (1998).
\bibitem{YY2} G. ’t Hooft,
{\em arXiv preprint} gr-qc/9310026 (1993).
\bibitem{27} S. S. Gubser, I. R. Klebanov, and A. M. Polyakov, 
{\em Physics Letters B}, \textbf{428}(1-2), 105–114 (1998).
\bibitem{28} G. W. Gibbons and S. W. Hawking, 
{\em Physical Review D}, \textbf{15}(10), 2752 (1977).
\bibitem{29} M. B. Ahmed, W. Cong, D. Kubizňák, R. B. Mann, and M. R. Visser, 
{\em Physical Review Letters}, \textbf{130}(18), 181401 (2023).

\bibitem{y1} Z. Gao, X. Kong, and L. Zhao,
{\em The European Physical Journal C}, \textbf{82}, 112 (2022).
\bibitem{y2} J. Sadeghi, M. A. S. Afshar, S. N. Gashti, and M. R. Alipour,
{\em Annals of Physics}, \textbf{460}, 169569 (2024).
\bibitem{y3} J. Sadeghi, M. Shokri, S. N. Gashti, and M. R. Alipour,
{\em General Relativity and Gravitation},
 \textbf{54}, 129 (2022).,
\bibitem{y4} M. R. Alipour, J. Sadeghi, and M. Shokri,
{\em Nuclear Physics B}, 990, 116184 (2023).
\bibitem{y5} T. Wang and L. Zhao,
{\em Physics Letters B}, \textbf{827}, 136935 (2022).

\bibitem{zz2} O. Aharony, S. S. Gubser, J. Maldacena, H. Ooguri, and Y. Oz,
{\em Physics Reports}, \textbf{323}, 183–386 (2000).



\bibitem{qq} K. Hashimoto, W. Sasaki, and T. Sumimoto,
{\em Journal of High Energy Physics}, \textbf{2019}, 1–17 (2019).
\bibitem{qqq} O. Aharony, S. S. Gubser, J. Maldacena, H. Ooguri, and Y. Oz,
{\em Physics Reports}, \textbf{323}, 183–386 (2000).


\bibitem{31} R. B. Wang, S. J. Ma, L. You, Y. C. Tang, Y. H. Feng, X. R. Hu, and J. B. Deng, 
{\em The European Physical Journal C}, \textbf{84}(11), 1161 (2024).
\bibitem{32} C. Rovelli, 
{\em Physical Review Letters}, \textbf{77}(16), 3288 (1996).
\bibitem{33} K. A. Meissner, 
{\em Classical and Quantum Gravity}, \textbf{21}(22), 5245 (2004).
\bibitem{34} A. Almheiri, N. Engelhardt, D. Marolf, and H. Maxfield, 
{\em Journal of High Energy Physics}, \textbf{2019}(12), 1–47 (2019).

\bibitem{YA} S. Y. Lin, M. H. Yu, X. H. Ge, and L. J. Tian,  
{\em Physical Review D}, \textbf{110}(4), 046008 (2024).
\end{thebibliography}
\end{document}